\documentstyle[tighten,preprint,eqsecnum,prl,aps,epsf]{revtex}
\def\pt{\mbox{$p_T$}}

\def\avkt{$\langle k_T \rangle$}

\def\piz{$\pi^0$}
\def\pim{$\pi^-$}
\def\delphi{$\Delta\phi$}
\def\pout{$p_{\rm{out}}$}
\def\pt{$p_T$}
\def\kt{$k_T$}

\def\sqrts{$\sqrt{s}$}
\def\ycm{$y_{\rm{cm}}$}

\def\NLOkt2{${\langle {\rm k_T^2} \rangle}$}
\begin{document}
%
\draft
\preprint{FERMILAB-Pub-97/351-E}
%
%
%
\title{Evidence for 
Parton \kt\ Effects in High-\pt\ Particle Production}

%
\author{
L.~Apanasevich,$^{4}$
J.~Bacigalupi,$^{1}$
W.~Baker,$^{3}$
M.~Begel,$^{9}$
S.~Blusk,$^{8}$
C.~Bromberg,$^{4}$
P.~Chang,$^{5}$
B.~Choudhary,$^{2}$
W.~H.~Chung,$^{8}$
L.~de~Barbaro,$^{9}$
W.~DeSoi,$^{9}$
W.~D\l ugosz,$^{5}$
J.~Dunlea,$^{9}$
E.~Engels,~Jr.,$^{8}$
G.~Fanourakis,$^{9}$
T.~Ferbel,$^{9}$
J.~Ftacnik,$^{9}$
D.~Garelick,$^{5}$
G.~Ginther,$^{9}$
M.~Glaubman,$^{5}$
P.~Gutierrez,$^{6}$
K.~Hartman,$^{7}$
J.~Huston,$^{4}$
C.~Johnstone,$^{3}$
V.~Kapoor,$^{2}$
J.~Kuehler,$^{6}$
C.~Lirakis,$^{5}$
F.~Lobkowicz,$^{9}$
P.~Lukens,$^{3}$
S.~Mani,$^{1}$
J.~Mansour,$^{9}$
A.~Maul,$^{4}$
R.~Miller,$^{4}$
B.~Y.~Oh,$^{7}$
G.~Osborne,$^{9}$
D.~Pellett,$^{1}$
E.~Prebys,$^{9}$
R.~Roser,$^{9}$
P.~Shepard,$^{8}$
R.~Shivpuri,$^{2}$
D.~Skow,$^{3}$
P.~Slattery,$^{9}$
L.~Sorrell,$^{4}$
D.~Striley,$^{5}$
W.~Toothacker,$^{7}$
N.~Varelas,$^{9}$
D.~Weerasundara,$^{8}$
J.~J.~Whitmore,$^{7}$
T.~Yasuda,$^{5}$
C.~Yosef,$^{4}$
M.~Zieli\'{n}ski,$^{9}$
V.~Zutshi$^{2}$
\\
{~}\\
\centerline{(Fermilab E706 Collaboration)}
{~}\\
}
\address{
\centerline{$^{1}$University of California-Davis, Davis, California 95616}
\centerline{$^{2}$University of Delhi, Delhi, India 110007}
\centerline{$^{3}$Fermi National Accelerator Laboratory, Batavia,
                   Illinois 60510}
\centerline{$^{4}$Michigan State University, East Lansing, Michigan 48824}
\centerline{$^{5}$Northeastern University, Boston, Massachusetts  02115}
\centerline{$^{6}$University of Oklahoma, Norman, Oklahoma  73019}
\centerline{$^{7}$Pennsylvania State University, University Park,
		   Pennsylvania 16802}
\centerline{$^{8}$University of Pittsburgh, Pittsburgh, 
Pennsylvania 15260}
\centerline{$^{9}$University of Rochester, Rochester, New York 14627}
}
\date{\today}
\maketitle
\begin{abstract}
Inclusive \piz\ and direct-photon cross sections in the kinematic 
range 3.5$<p_T<$12~GeV/$c$ with central rapidities (\ycm) are 
presented for 530 and 800~GeV/$c$ proton beams and a 515~GeV/$c$ \pim\ 
beam incident on Be targets.
Current Next-to-Leading-Order perturbative QCD calculations fail to 
adequately describe the data for conventional choices of scales.
Kinematic distributions from these hard scattering events provide evidence 
that the interacting partons carry significant initial-state parton 
transverse momentum (\kt). 
Incorporating these \kt\ effects phenomenologically greatly improves the 
agreement between calculations and the measured cross sections.
\end{abstract}

\pacs{PACS numbers: 12.38.Qk, 13.85.Qk, 13.85.Ni, 14.20.Dh}
%
%
%

In recent years, perturbative QCD (PQCD)
has been tested in a wide variety of processes involving 
strong interactions at short distances, and
increasing attention is now being directed towards areas that may be sensitive
to shortcomings in the current theoretical description~\cite{curpqcd}.
The high statistics samples of hard scattering data accumulated
by Fermilab fixed-target experiment E706 provide an opportunity to probe 
such issues.
This paper presents comparisons of 
PQCD calculations to our data on the production
of direct photons and \piz's with large transverse momenta (\pt).
Direct-photon data have long been 
expected to provide an accurate determination of the distributions of 
gluons in hadrons, especially at large 
longitudinal momentum fraction ($x$), 
where information has proven difficult to obtain from other measurements.
Inclusive meson production at large \pt\ probes a 
different mix of hard scattering processes and provides insights into 
parton fragmentation. 
For conventional choices of parameters,
our data are not described satisfactorily by 
Next-to-Leading-Order (NLO) PQCD calculations~\cite{nlodpa}.
Resolving the observed discrepancies is important for improving the 
understanding of both parton distribution functions (PDF) 
and parton fragmentation functions (FF).

Several interesting aspects of QCD contributions 
beyond Leading-Order (LO) can be investigated 
experimentally through studies of processes sensitive to 
transverse motion of the partons prior to the hard scatter. 
This \kt\ is presumably due to effects of hadron size (primordial \kt)
as well as initial-state gluon radiation.
Measurements of Drell-Yan pair production~\cite{dykt} 
and direct di-photon production~\cite{otherkt} have
demonstrated the presence of substantial effective \kt,
(significantly larger than can be attributed to primordial \kt),
and have revealed a significant \sqrts\ dependence of \avkt.  
A resummation of soft gluon emissions has recently been used to 
reproduce the size of the effect observed in the WA70  
direct di-photon data~\cite{fergani}.
Other data also suggest 
\avkt\ values larger than those expected from NLO PQCD calculations. 
Recent comparisons of \pt\
spectra from charm-particle hadroproduction to NLO PQCD results 
provide evidence that supplemental \kt\ may be required to
properly describe the data~\cite{frixione}. 
Likewise, it has been suggested that
the observed pattern of discrepancies between data from various 
direct-photon experiments and results from NLO PQCD calculations 
could be related to \kt\ effects~\cite{cteqkt}.

E706 is designed to measure the production of direct photons, neutral 
mesons, and associated particles at high \pt. The apparatus features a 
large lead and liquid argon electromagnetic calorimeter and a charged 
particle spectrometer~\cite{D90}.
The experiment accumulated 
$\approx$10~events/pb of \pim\ beam data at 515~GeV/$c$, 
$\approx$9~events/pb of proton beam data at 530~GeV/$c$, and 
$\approx$11~events/pb of proton beam data at 800~GeV/$c$ 
on Be, Cu, and H targets (primarily Be).
A variety of event selection triggers sensitive to high-\pt\
electromagnetic showers were employed (using different
prescale factors) to accumulate data over a broad range of \pt.

The steep \pt\ dependences of neutral meson and 
direct-photon production make the measured cross sections very
sensitive to uncertainties in the energy calibration.
Therefore, achieving a precise and accurate calibration of the response of the
electromagnetic calorimeter was essential to the success of E706.  
As a result of detailed studies of the data acquired, the uncertainty in 
the calibration of the energy response of the calorimeter was reduced to 
less than 0.5\%~\cite{lac_cal}.

The single-photon sample is composed of those photons not identified as 
elements of reconstructed two photon decays of \piz\ or $\eta$ 
meson candidates. 
The direct-photon signal is extracted from the single-photon sample
via statistical subtraction of the background contributions.
These backgrounds are primarily due to photons from
unreconstructed decays of neutral mesons.
Failure to correctly identify a photon as originating from 
a \piz\ or $\eta$ decay 
occurs when the other photon from that decay converts
in the target region, escapes the fiducial volume of the calorimeter, or
is otherwise not reconstructed.
Sources of direct-photon background have been modeled using the {\sc herwig} 
event generator~\cite{herwig} and a detailed {\sc geant} 
simulation~\cite{geant3} of the spectrometer response.  
These Monte Carlo generated events have been weighted to accurately
represent our measured neutral meson production spectra.

\begin{figure}
\vskip -0.8 truecm
\epsfxsize=6.5 truein
\centerline{\epsffile[0 72 612 720]{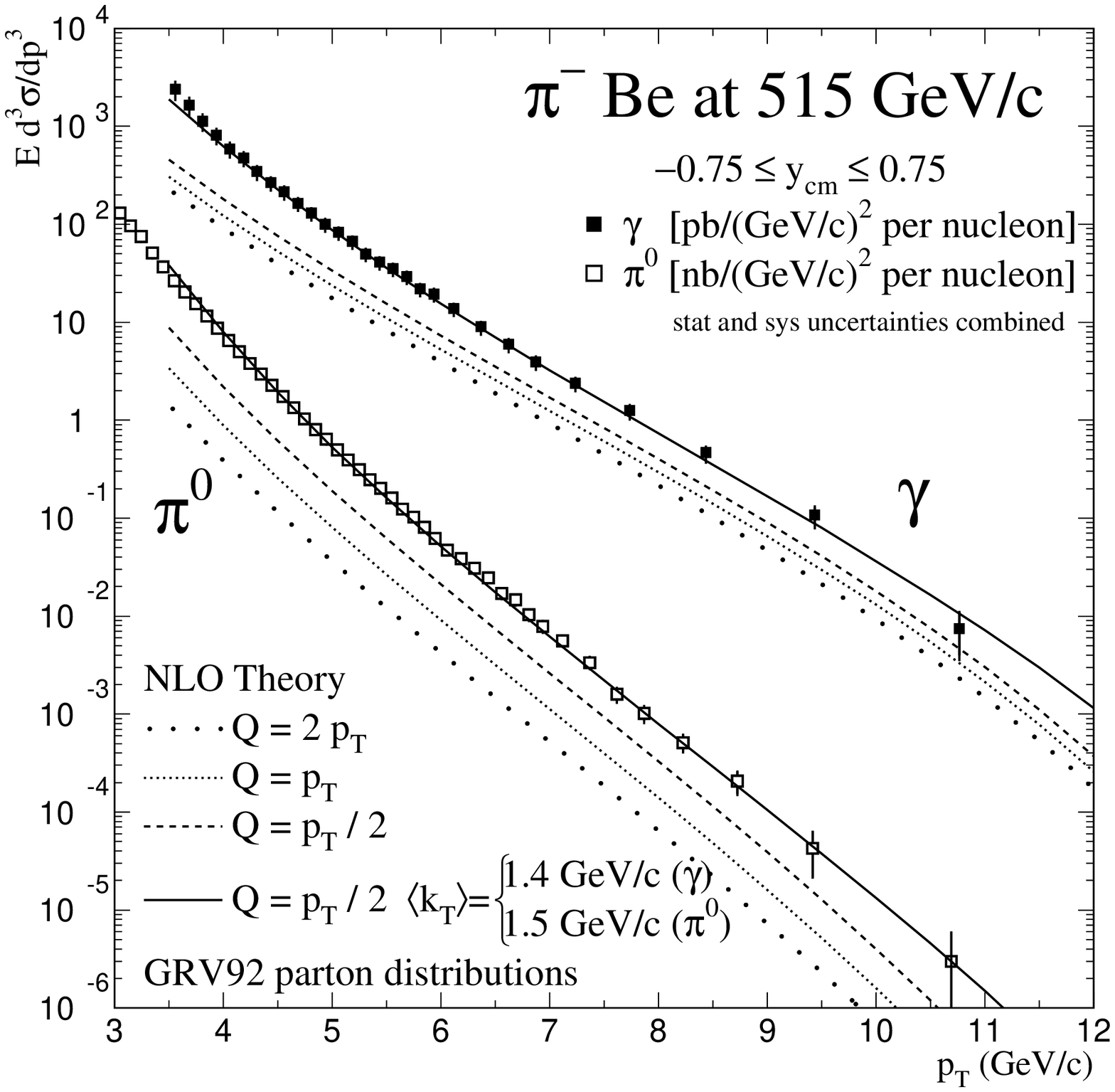}}
\vskip -1.4 truecm
\caption{The \piz\ and direct-photon inclusive cross sections
as functions of \pt\ for 515~GeV/$c$ \pim-nucleon interactions 
compared to NLO PQCD results 
for several choices of 
scales.
The solid curves show the NLO PQCD results for $Q=p_{T}/2$ scales
adjusted for supplemental \avkt.
(Note that the units for the \piz\ and $\gamma$ results 
differ by a factor of 1000.)} 
\label{fig:xs_qsq_nlo_515}
\end{figure}

\begin{figure}
\vskip -0.8 truecm
\epsfxsize=6.5 truein
\centerline{\epsffile[0 72 612 720]{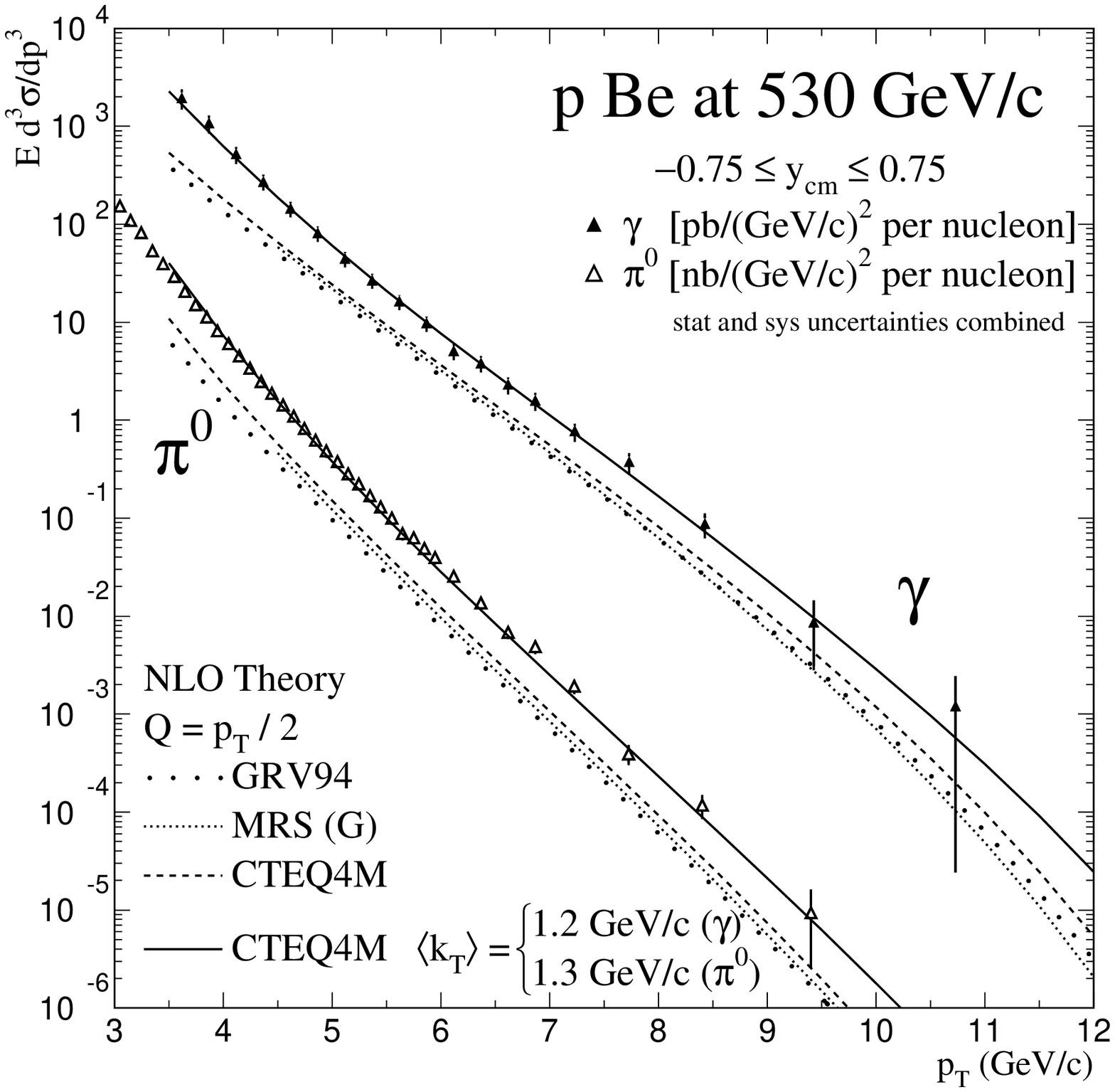}}
\vskip -1.4 truecm
\caption{The \piz\ and direct-photon inclusive cross sections
as functions of \pt\ for 530~GeV/$c$ proton-nucleon interactions
compared to NLO PQCD results for several
choices of PDF.
The solid curves show the NLO result (using the CTEQ4M PDF)
adjusted for supplemental \avkt.
(Note that the units for the \piz\ and $\gamma$ results 
differ by a factor of 1000.)} 
\label{fig:xs_pdf_530}
\end{figure}

\begin{figure}
\vskip -0.8 truecm
\epsfxsize=6.5 truein
\centerline{\epsffile[0 72 612 720]{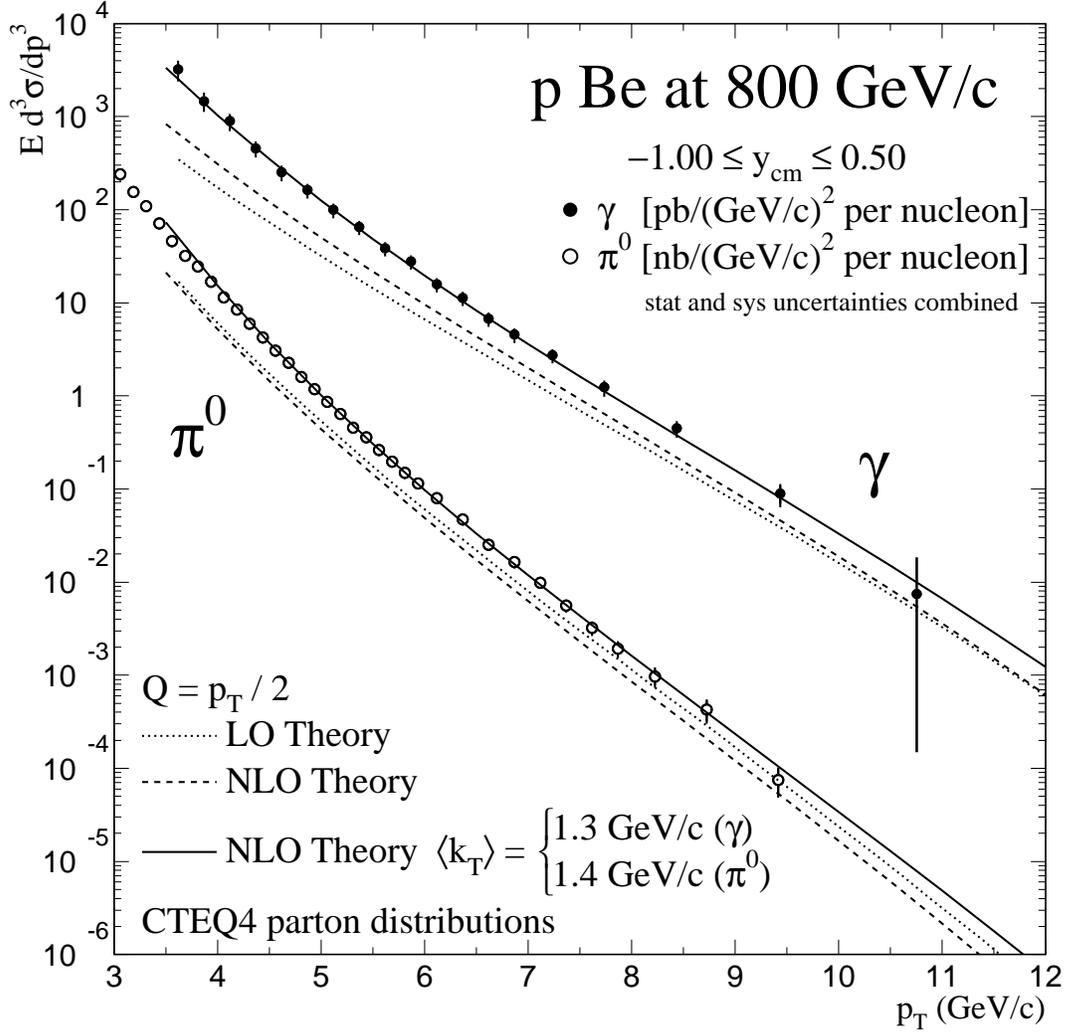}}
\vskip -1.4 truecm
\caption{The \piz\ and direct-photon 
cross sections
as functions of \pt\ for 800~GeV/$c$ proton-nucleon interactions
compared to LO and NLO PQCD results.
The solid curves show NLO results
adjusted for supplemental \avkt.
(Note that the units for the \piz\ and $\gamma$ results differ by 
a factor of 1000.)} 
\label{fig:xs_nlo_lo_800}
\end{figure}

The \pt\ dependences of inclusive \piz\ and direct-photon cross 
sections are shown in Figs.~\ref{fig:xs_qsq_nlo_515}, 
\ref{fig:xs_pdf_530}, and \ref{fig:xs_nlo_lo_800}.  
The results of NLO PQCD calculations
(using BKK FF for the \piz~\cite{bkk}) are compared with the data~\cite{acor}.
For simplicity, all QCD scales (renormalization, factorization, and, 
where appropriate, fragmentation) have been set equal.
The broken curves in Fig.~\ref{fig:xs_qsq_nlo_515} represent the
results of NLO PQCD calculations using conventional choices of
scales (and GRV PDF~\cite{grv92}).  
The calculations are quite sensitive to the scales
(an indication of the importance of still higher order contributions),
but even for rather small scales, the NLO calculations 
fail to describe our data.
Using other recent PDF~\cite{grv94+mrsg,cteq4} in the calculations 
also does not adequately account for the 
discrepancy between the NLO PQCD results and our data 
(broken curves in Fig.~\ref{fig:xs_pdf_530})~\cite{pdfsens}.
Differences between LO~\cite{owens}
and NLO calculations are likewise not large
when compared to the difference observed between either
and the measured cross sections (Fig.~\ref{fig:xs_nlo_lo_800}).  
However, PQCD at NLO
may not adequately account for soft gluon radiation that imparts
an effective transverse momentum to the incident partons.

\begin{figure}
\vskip -0.8 truecm
\epsfxsize=6.5 truein
\centerline{\epsffile[0 72 612 720]{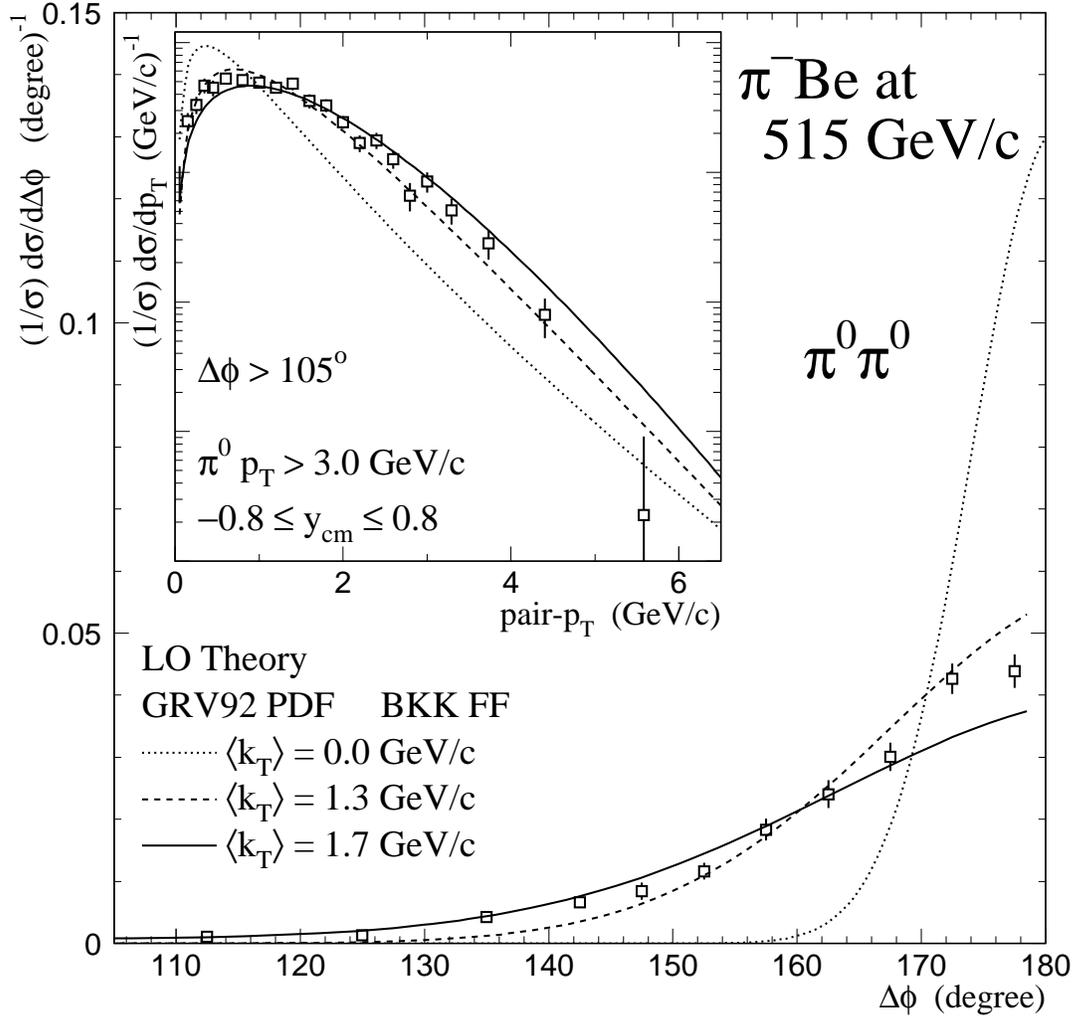}}
\vskip -1.4 truecm
\caption{The \delphi\ distribution for high-mass \piz\ pairs
produced in 515~GeV/$c$ \pim-nucleon interactions compared to
curves showing LO PQCD results using various \avkt\ values.
The inset shows the pair-\pt\ distribution for such pairs
and the corresponding results of LO PQCD calculations.
}
\label{fig:mkt}
\end{figure}

Kinematic distributions for high-mass pairs of particles
directly probe the transverse momentum of incident partons in
hard scattering events.
The \pt\ distribution of \piz\ pairs
produced in 515~GeV/$c$ \pim-nucleon collisions is shown in the inset of
Fig.~\ref{fig:mkt}. 
The results of LO PQCD
calculations in which the incident partons have
Gaussian transverse momentum distributions with 
\avkt\ =~1.3~GeV/$c$ (dashed curve) and \avkt\ =~1.7~GeV/$c$ (solid curve)
are shown in Fig.~\ref{fig:mkt}~\cite{owens};
the dotted curve does not include any incident parton \kt.
For these high-mass \piz\ pairs, Fig.~\ref{fig:mkt} also shows
the distribution of the angle between the two \piz's 
in the transverse plane (\delphi), and provides a comparison to
LO PQCD, both with and without supplemental \kt.
Fragmentation itself generates significant pair-\pt\ and contributes
to the width of the \delphi\ distributions, 
but supplemental \avkt\ $>$~1~GeV/$c$ provides a
much better description of the data.

\begin{figure}
\vskip -0.8 truecm
\epsfxsize=6.5 truein
\centerline{\epsffile[0 72 612 720]{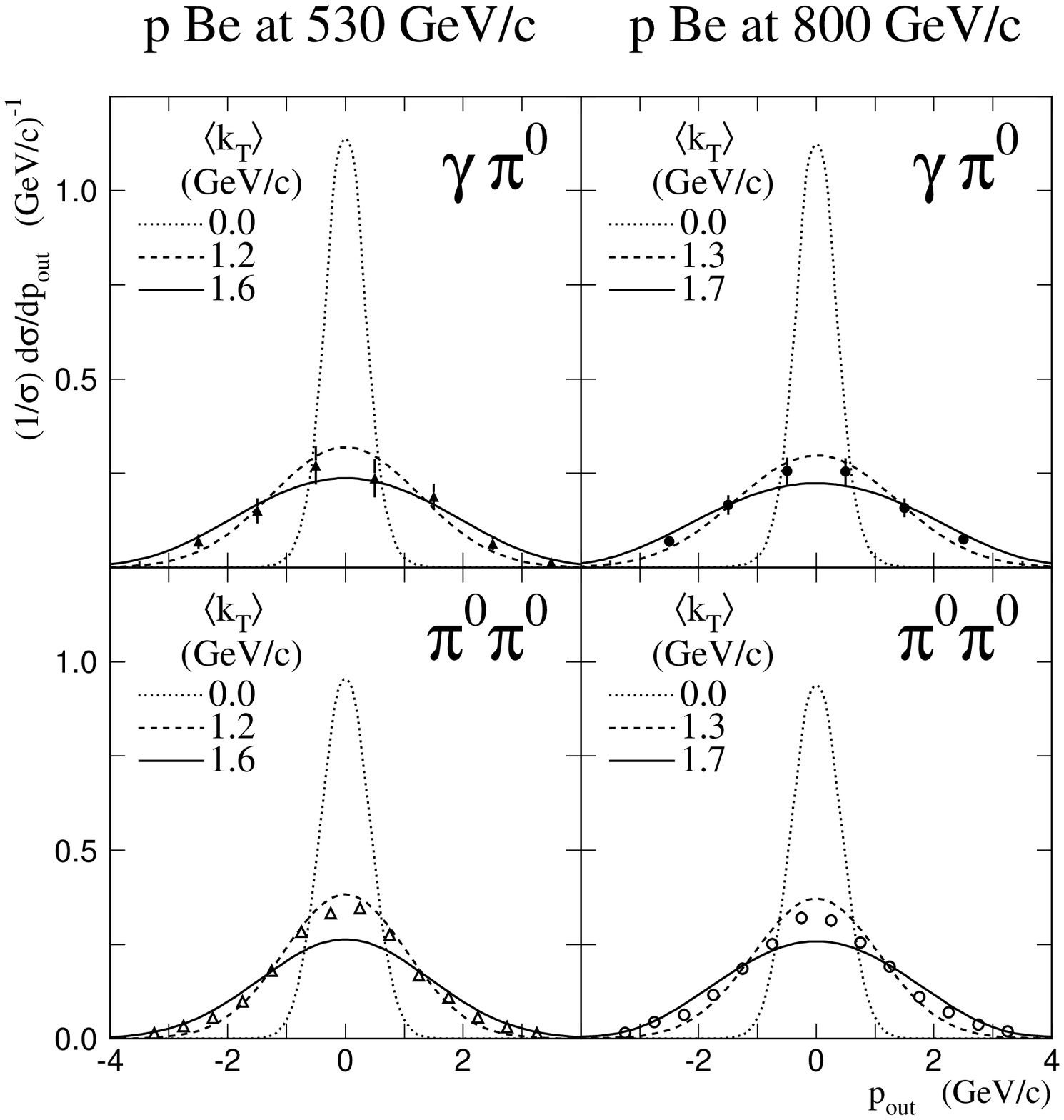}}
\vskip -1.2 truecm
\caption{
The out-of-plane momentum distributions for high-mass pairs  
produced in proton-nucleon interactions at 530 and 800 GeV/$c$ 
compared to results of LO PQCD calculations 
(using CTEQ4L PDF) for several \avkt\ values.
}
\label{fig:mpout}
\end{figure}

Another kinematic variable that is sensitive to \kt\ 
is the out-of-plane momentum, \pout\  
(the component of the momentum of one high-\pt\ particle, perpendicular 
to the plane defined by the incident beam direction and the direction of 
the other high-\pt\ particle).   
The \pout\ distributions for $\gamma$\piz\ pairs produced 
in proton-nucleon interactions are shown in the upper part of
Fig.~\ref{fig:mpout}, compared to LO results without \kt, and 
with \avkt\ values chosen to bracket the data.
The \pout\ distributions for  \piz\ pairs are also shown
in Fig.~\ref{fig:mpout} (lower plots).  
These distributions (pair-\pt, \delphi, and \pout) show clear 
evidence for the presence of significant \kt\ ($>$~1~GeV/$c$) 
in the hard scattering 
interactions.  The corresponding distributions for our other data
samples also support this conclusion~\cite{begel}.

Our preliminary analyses of the kinematic distributions of pairs of 
direct photons, as well as studies of the distribution of the fractional
momentum carried by individual charged particles in jets recoiling 
against isolated photons, also show evidence of substantial \kt,
as do our comparisons of the measured high-\pt\ charged-$D$ cross section 
to NLO PQCD results~\cite{D90}.  All these results suggest a
supplemental \avkt\ of order 1~GeV/$c$.

Since the inclusive spectra fall rapidly with increasing \pt, 
the introduction of \kt-smearing has a significant effect on predicted 
cross sections.  
To approximate the effect of supplemental \kt\ smearing on the inclusive 
NLO PQCD calculations for direct-photon (and \piz) production, we  
calculated \kt\ factors (as functions of \pt) for different 
values of \avkt, by computing 
ratios of results from LO PQCD calculations~\cite{owens} 
for different \avkt\ values
compared to results without \kt~\cite{ktscale}.
These same \kt\ factors were then applied to the results of 
NLO PQCD calculations.
As indicated by the solid curves in Figs.~\ref{fig:xs_qsq_nlo_515}, 
\ref{fig:xs_pdf_530}, and \ref{fig:xs_nlo_lo_800}, 
reasonable representations of both the direct-photon and \piz\ results 
are obtained using \avkt\ values $> 1$~GeV/$c$~\cite{llkt}.
The kinematic distributions exhibit a pattern
consistent with increasing \avkt\ as \sqrts\ increases,  a trend
reflected in the choices of \avkt\ factors employed in the
theory curves (solid curves) shown in the inclusive cross section plots.

\begin{figure}
\vskip -0.8 truecm
\epsfxsize=6.5 truein
\centerline{\epsffile[0 72 612 720]{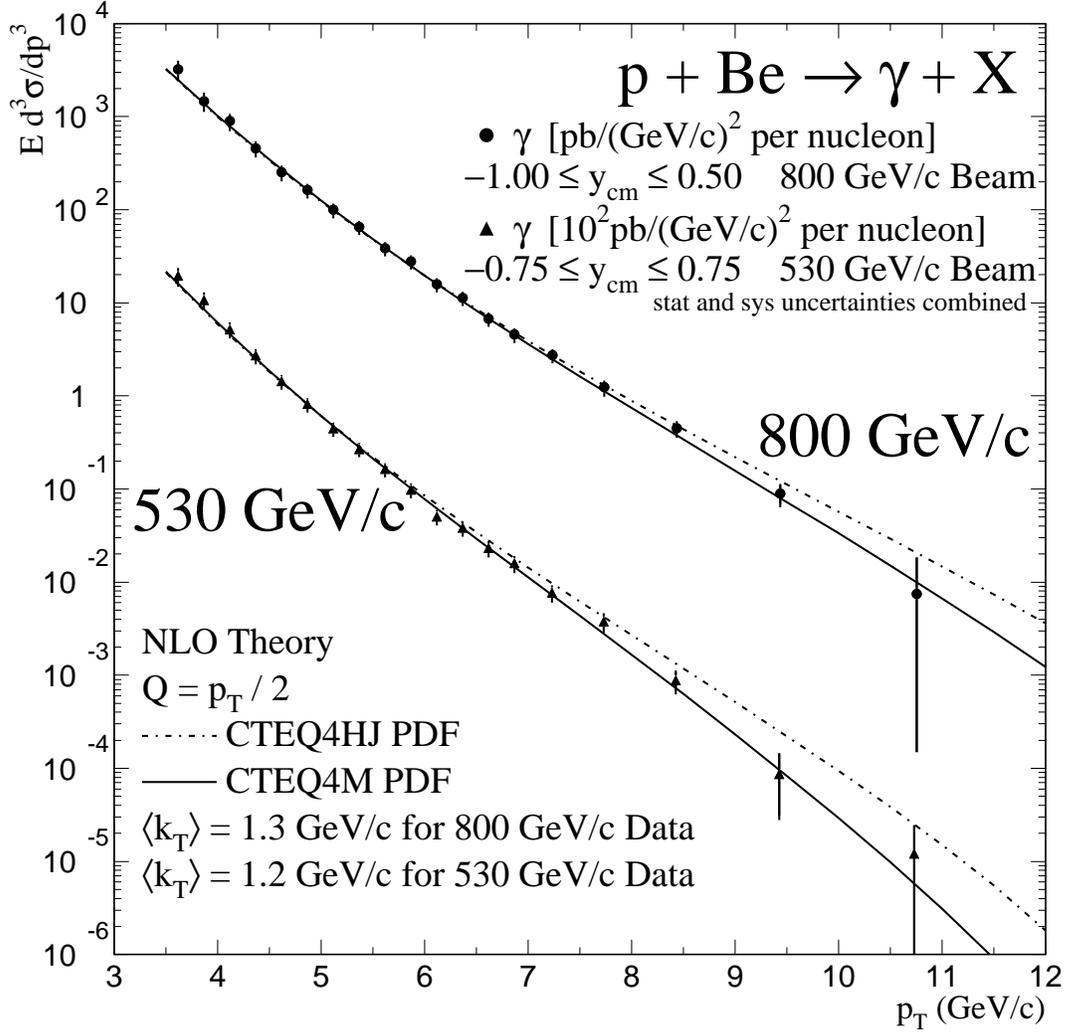}}
\vskip -1.4 truecm
\caption{Direct-photon inclusive cross sections as functions
of \pt\ for 530 and 800~GeV/$c$ proton-nucleon interactions
compared to results of NLO PQCD calculations
using CTEQ4HJ (dot-dashed curve) and CTEQ4M (solid curve) PDF.
Factors for supplemental \avkt\ are included.
(Note that the units for the 530 and 800 GeV/$c$
results differ by a factor of 100.)}
\label{fig:xs_kt_hj}
\end{figure}

As an illustration of the sensitivity of our data to the gluon
distribution, 
Fig.~\ref{fig:xs_kt_hj} compares 
our direct-photon cross sections to NLO PQCD calculations using CTEQ4M
and CTEQ4HJ PDF~\cite{cteq4}.
Once soft-gluon effects are satisfactorily taken into account, 
either approximately
as in this paper or in a more theoretically rigorous manner, 
our data can be used to help discriminate between PDF that otherwise 
provide acceptable descriptions of the data sets used in Ref.~\cite{cteq4}.

In conclusion, we have measured the inclusive production of high-\pt\
neutral mesons and direct photons by 530~GeV/$c$ 
and 800~GeV/$c$ proton and 515~GeV/$c$ \pim\ beams.
Current NLO PQCD calculations (which exhibit substantial dependences 
on QCD scales) fail to account for the measured cross sections using 
conventional choices of scales.  
Significant \kt\ effects ($>$~1~GeV/$c$) have been observed in the 
kinematic distributions of high-mass pairs of \piz's, as well as 
high-mass $\gamma$\piz\ pairs.  
A simple implementation of supplemental parton \kt\ in
PQCD calculations, using \kt\ values consistent with observations, 
provides a reasonable description of the inclusive cross sections.
Our high statistics 
direct-photon data samples are directly sensitive to the gluon
distribution at large $x$ values.  An improved theoretical
understanding of soft-gluon effects in inclusive
direct-photon production will facilitate the global determination
of the gluon distribution function.  

We are pleased to acknowledge the many contributions of the Fermilab 
staff.
This research has been supported in part by the U.~S. Department of Energy, 
the National Science Foundation, and the Universities Grants Commission of 
India.
%
%

\begin{references}
\bibitem{acor}
Each theory calculation presented in this paper has been adjusted to account
  for nuclear effects. Our conclusions are fully supported by our hydrogen
  target data samples. For the inclusive high-\pt\ \piz\ and direct-photon
  calculations, theory results were adjusted by $A^{\alpha-1}$, assuming
  $\alpha=1.08$ and $\alpha=1.04$, respectively.
\bibitem{pdfsens}
The variation in these results does not reflect the full uncertainty in the
  knowledge of the gluon distribution at large $x$ values since little data
  have been available in this domain, and there is substantial overlap in the
  data used and the techniques employed by different groups in evaluating the
  various PDF.
\bibitem{ktscale}
A possible variation of \avkt\ with \pt, expected in PQCD, has been neglected
  in this analysis.
\bibitem{llkt}
The \avkt\ values that provide a corresponding level of agreement between LO
  direct-photon calculations (not shown) and our inclusive cross sections are
  somewhat larger ($\approx$~10\%) than the corresponding values for the NLO
  comparisons.
\end{references}

%
%
%
%
\newpage
\noindent
%
\end{document}